\documentclass[twoside, twocolumn, a4paper]{article} %twocolumn

\usepackage{lastpage}
\usepackage{amsmath,amsfonts,amssymb,amsthm,epsfig,epstopdf,titling,url,array}
\theoremstyle{definition}

\usepackage{datetime}
\usepackage{tikz}

\usepackage{tabularx,booktabs}
\usepackage{fontawesome5}

\usepackage[draft]{hyperref}
\usepackage[toc,style=altlistgroup, hyperfirst=false]{glossaries}

\usepackage[sc]{mathpazo} % Use the Palatino font
\usepackage[T1]{fontenc} % Use 8-bit encoding that has 256 glyphs
\usepackage{microtype} % Slightly tweak font spacing for aesthetics
\usepackage[utf8x]{inputenc}
\usepackage[english]{babel} % Language hyphenation and typographical rules
\usepackage{import}
\usepackage{graphicx}
\usepackage{xr}

\usepackage[hmarginratio=1:1,top=32mm,columnsep=20pt]{geometry} % Document margins
\usepackage[hang, small,labelfont=bf,up,textfont=it,up]{caption} % Custom captions under/above floats in tables or figures
\usepackage{booktabs} % Horizontal rules in tables

\usepackage{enumitem} % Customized lists
\setlist[itemize]{noitemsep} % Make itemize lists more compact

\usepackage{abstract} % Allows abstract customization
\usepackage{titlesec} % Allows customization of titles
\renewcommand\thesection{\Roman{section}} % Roman numerals for the sections
\renewcommand\thesubsection{\roman{subsection}} % roman numerals for subsections

\titleformat{\section}[block]{\large\scshape\centering}{\thesection.}{1em}{} % Change the look of the section titles
\titleformat{\subsection}[block]{\large}{\thesubsection.}{1em}{} % Change the look of the section titles
\newcommand{\articleTitle}{Argentum: a collaborative saving and investment platform for unstable countries}
\usepackage{fancyhdr} % Headers and footers
\usepackage{titling} % Customizing the title section

\usepackage{hyperref} % For hyperlinks in the PDF

\pretitle{\begin{center}\Huge\bfseries} % Article title formatting
\posttitle{\end{center}} % Article title closing formatting
\title{\articleTitle{}} % Article title
\author{
  \textsc{Leonardo Javier Belén}, \textsc{Alejandro Baranek}, \textsc{Xavier Ignacio González} \\ % Your name
  \normalsize \href{mailto:leobelen@gmail.com}{leobelen@gmail.com}, \href{abaranek@dc.uba.ar}{abaranek@dc.uba.ar}, \href{xavierign@gmail.com}{xavierign@gmail.com}
}

\date{\today}
\renewcommand{\maketitlehookd}{%
\begin{abstract}
\noindent A crypto coin designed to provide a stabilization instrument backed up by minded like financial investments instruments to maintain the purchase value of savings across time, in order to construct new tools for unstable economies.
\end{abstract}
}

\makeindex
\makeglossaries

\usepackage{Sweave}
\begin{document}
\Sconcordance{concordance:argentum.tex:argentum.Rnw:%
1 88 1 1 0 7 1}
\Sconcordance{concordance:argentum.tex:./include/introduction.Rnw:ofs 97:%
1 14 1}
\Sconcordance{concordance:argentum.tex:./include/concept.Rnw:ofs 112:%
1 44 1}
\Sconcordance{concordance:argentum.tex:./include/details.Rnw:ofs 157:%
1 134 1}
\Sconcordance{concordance:argentum.tex:./include/associatedProducts.Rnw:ofs 292:%
1 5 1}
\Sconcordance{concordance:argentum.tex:./include/aurum.Rnw:ofs 298:%
1 30 1}
\Sconcordance{concordance:argentum.tex:./include/monthlyBidding.Rnw:ofs 329:%
1 21 1}
\Sconcordance{concordance:argentum.tex:./include/platinum.Rnw:ofs 351:%
1 15 1}
\Sconcordance{concordance:argentum.tex:./include/benefits.Rnw:ofs 367:%
1 10 1}
\Sconcordance{concordance:argentum.tex:./include/conclusion.Rnw:ofs 378:%
1 14 1}
\Sconcordance{concordance:argentum.tex:./include/anexes.Rnw:ofs 393:%
1 84 1}
\Sconcordance{concordance:argentum.tex:argentum.Rnw:ofs 478:%
104 8 1}

\setkeys{Gin}{width=\linewidth} %this sets up the width of Sweave generated plots.
\setlength{\parskip}{1ex plus 0.5ex minus 0.2ex}
% Print the title
\maketitle

% -*- coding: utf-8 -*-
% !Rnw root = ../argentum.Rnw
%\section{TODO}

%\begin{itemize}
%  \item Replace ARS an USD with \$
%\end{itemize}

\section{Introduction}

Among the central and most debated issues regarding the economy of developing countries are inflation, which is a sustained and generalized increase in the price level of goods and services over a period of time\cite{inflation_wiki}, and hyperinflation, which is a very high and typically accelerating inflation\footnote{There is no consensus about the differences of inflation-high inflation and hyperinflation. This works focus on high inflation for Argentina's standards, that is more thant 20\% and less than 60\% -which is considered hyperinflation by some international organisms-}. Causes of inflation can be attributed to different interrelated factors that can include: demand supply gaps, increase in money supply, currency devaluation, increase in the prices of inputs (e.g. oil, labor costs), Economy credibility, market structure, and expectations. Inflation can turn to hyperinflation by a spiral mechanism in which one factor induces the other. One of the most important effects in inflation is a progressive decrease in the purchasing power of the currency. This harmful effect is not evenly distributed in society and generally afflicts more to lower segments. Other related consequences include: discouraging savings and investments, inefficiencies in goods and money markets, and implicit tax increase\cite{inflation_wiki}. Although much research has been done to control inflation, by 2018 more than 20 countries have projected inflation indexes over 10\%\cite{countries}. In these countries, inflation distort prices signals introducing uncertainity by eroding the confidence on the prices system.

As a big step towards mitigating some of the negative effects of inflation, we contribute \textbf{Argentum}, a collaborative platform and blockchain based cryptocurrency suitable for unstable countries. By its definition and structure, \textbf{Argentum} is aimed to provide to its users a stable currency in a volatile economy, making available a financial instrument for savers and investors to maintain the purchasing power over time. Backed up by financial investments strategies available in the host country, the token value will rely in provable assets instead of confidence on the token itself. \textbf{Argentum} serves as a stable value measure for goods and services and can serve as a medium of exchange -when enough relevance is achived-. \textbf{Argentum} is conceived not to be a form of enrichment through speculation, on the contrary, it is aimed to be a tool for the common citizen to protect himself against the uncertainity using tools previously available to specialists.

In the next section, \textbf{Argentum} concept is elaborated: the rules that regulate its operation and the principles that frame its philosophy are introduced. Section \ref{BF56EC92-1E45-4656-A440-CEA20D7066B5} elaborates about the system in which this concept is developed and finally, in Section \ref{A3CDF833-7291-4D84-9342-04E0A108DDA8}, some conclusions are discussed.
% -*- coding: utf-8 -*-
% !Rnw root = ../argentum.Rnw

\section{Concept}
\label{section:argentum_concept}

\textbf{Argentum} is a concept with dual meaning. It is a cryptocurrency based on blockchain and a technological platform integrated by a community of savers, investors, administrators, developers and partners. Savers and investors are the main roles in the community, \textbf{Argentum} cryptocurrency operation is governed by these 3 main rules:

\begin{enumerate}
  \item  \textbf{Saving Rule}: \textbf{Argentum} valuation is pegged to the Argentine Consumer Price Index (IPC es-CPI en)\cite{Wikipedia_IPC_Argentina}, which is the recommended  inflation proxy in the world\cite{cpi_ilo_manual,cpi_manual_updates} for household expenditures. So, \textbf{Argentum} will be a constant (deflected) valuation measure for Argentina's goods and services, which is considered useful in countries with high inflation\footnote{In Argentina, for example, real state and capital assets are commonly traded in $U\$D$.}. %Along with this, \textbf{Argentum}  is intended to pay a moderate interest on the currency, around 2\%, which the user will see as an incoming transaction to his/her account.

  \item \textbf{Creation Rule}: Users can acquire and use \textbf{Argentum} freely as long as they keep their participation under a safety margin, to be defined dinamically by the  \textbf{Argentum Inc.}. As all  \textbf{Argentum} tokens have to be based on the opportunities that can be found to back those amounts, it may be possible to for the governing entity to accept the amounts, put them on hold for a period of time less than a month, and if no opportunities have been found, return the amounts to the buyer.

  \item \textbf{Selling Rule}: Although \textbf{Argentum} holders can transfer amounts unrestrictedly to other account holders, selling (alas, obtaining other currencies in exchange for their \textbf{Argentum} amounts), can be done only if the total amount sold for a given period of time does not compromise the estability of the  other  \textbf{Argentum} holders. That threshold will be defined dinamically by the  \textbf{Argentum Inc.}. This will discourage the use of \textbf{Argentum}for speculative proposes.

\end{enumerate}

In order to make possible to implement the previous rules, \textbf{Argentum} aims to create a new, standards based, state-of-the-art technology based currency with following principles:

\begin{enumerate}

  \item \textbf{Purchasing power}:In order to maintain the purchasing power, it will pegged to each currency plus the Consumer Price Index, which in the case of Argentina is called IPC\cite{Wikipedia_IPC_Argentina}. In other words, \textbf{Argentum} is a stablecoin\cite{Forbes_explaining_stable_coins} that during economic hardships maintains the purchasing power and during stable times is simply pegged to the currency, ensuring stability in the long term.

  \item \textbf{Risk Minimization}: the coin is designed to deal with multiple scenarios, but unlike most of the tokens around, meant to provide earnings maximization in the short term (hence increasing the volatility), \textbf{Argentum} will be backed up by a team of Financial Operators with the express intent of maintaining the value over time, allowing its use as a safe earning method, including measures such as:
  \begin{enumerate}
      \item Actively changing the number of tokens in the market to equalize its value to the targeted fiat value.
      \item Making conservative investments on each economy it is operating to match the targeted earnings rate, making heavy use of optimized portfolios with cyclic and contra-ciclyc assets.
      \item Adding restrictions to pull outs to avoid economic surges over time, based on the amount of liquid assets. However, this restriction will not apply to transfers between \textbf{Argentum} users.
   \end{enumerate}
  \item \textbf{Transparency and Traceability}: In order to guarantee traceability \textbf{Argentum} will:
  \begin{enumerate}
    \item use blockchain based technologies, an already proven technology to store in a collaborative manner all transactions, making it nearly impossible to counterfeit
    \item be open source, to ensure community help and audit.
    \item enforce non-anonymity, which aims to identify every wallet with a uniquely identified user, making impossible to hide assets on Argentum, in the same sense that bank accounts cannot be anonymous. In this regard, the user information will not be held in plain text but encrypted and accesible only on specific situations by the jurisdictional authorities.
    \item make periodic publication of progress, which is intended to inform \textbf{Argentum} holders on the overall progress of the project.
    \item publish online publication of the exchange rate, with the express aim of mitigate speculation on the currency's price.
  \end{enumerate}
  \item \textbf{Algoritmic focalization}: Every human is a world of decisions. As such \textbf{Argentum} is set to leverage the full power of algorithms to maximize its objectives with the minimal impact, allowing it to run at all times, safely and based on known rules and procedures. Initially, the philosophy behind the elegibility of a trading algorithm is to exploit distortions (or arbitrate) in relative prices, a known inflation effect\cite{Costs_of_inflation, RePEc:san:cdmawp:0611}.
  \item \textbf{Ecosystem friendly}: \textbf{Argentum} is meant to be the cornerstone of many other projects that need to be built upon a realiable currency.
  \item \textbf{Non-anonimity}: All holders have to be registered properly before being able to operate in the system, and no hidden or unnamed accounts are possible, both to increase the confidence between holders and to comply with government regulations. Information used to identify holders are not meant to be on the blockchain and/or visible from outside of \textbf{Argentum inc.}.
\end{enumerate}

% -*- coding: utf-8 -*-
% !Rnw root = ../argentum.Rnw

\section{Details}
\label{BF56EC92-1E45-4656-A440-CEA20D7066B5}
\subsection{Brief analysis of Incentives}
\label{section:Brief_analysis_of_incentives}
The proposed system heavily relies on incentives both to educate and reward users in order to guide them into good saving and investment practices by mantaining the value across time and boundaries.

With these incentives,  \textbf{Argentum} intends to align the user with this project, making a virtuous circle of growth and cooperation.

It stands to reason that it is expectable that both savers and investors will keep their investments at least up to the premium rate as to earn the premium in full. On the same manner, the premium procedure is transparent to the user and  \textbf{Argentum Inc.} at all times, adding clarity to the operations, and prevision on what amount is compromised for premiums and which is available to  \textbf{Argentum Inc.} as unrestricted funds.

\textbf{Argentum Inc.} will publish on a daily basis a dashboard with information to allow users to follow up the progress of the project.

%  {\color{red} Con estos incentivos, se espera que los usuarios mantengan el dinero argentum para cobrar la mayor proporcion posible del premio. A su vez, la regla permite valuar que porcentaje de las ganancias estan comprometidas a usuarios y que porcentaje es de libre disponibilidad, para el fin que Argentum Inc. determine}
.

\subsection{Accounting model}
\textbf{Argentum} has an incentive model that includes a monthly interest rate and a bonus premium 10 semesters after deposit which will be paid on \textbf{Argentum} if it is in expansion  and in $AR\$$ if it is in contraction.

In order to understand \textbf{Argentum}'s accounting model there are two main accounts: 1) tokens in the public domain and 2) Withheld tokens or tokens that are generated by \textbf{Argentum Inc.}. The withheld tokens are generated on earnings while the net investment position yields earnings and are destroyed if the net investement position to isolate the impact of investments on \textbf{Argentum}'s price. Additionally, a small amount of withheld tokens will be daily deactivated to peg \textbf{Argentum}'s price to the inflation rate.

The withheld account includes also the account from which are sourced interest and premium payments to the users. Additionally, the withheld account includes a anti cyclic fund, which is funded by a ratio over the earnings to cover for contractionary period interest and premiums payments. Those funds could be used on riskier investments.

The tokens on the public domain account is composed, in turn, by the tokens that are owned by the public and the tokens that are being offered to the public but have not been adjudicated yet.

It has to be noted that each user has - at least - two accounts, one in \textbf{Argentum} and one in $AR\$$. While the \textbf{Argentum} account will be used on the daily basis, the  $AR\$$ account will be used to capitalize during the bidding process and to refund and award amounts in  $AR\$$ by the system. All amounts in  $AR\$$ will be transferable to the financial system or used in the bidding process at any time. As the philosophy entails transparency, both savings and investment funds into the project have to be originated on recognized financial institutions.

\begin{figure}
    \centering
    \includegraphics[width=\linewidth]{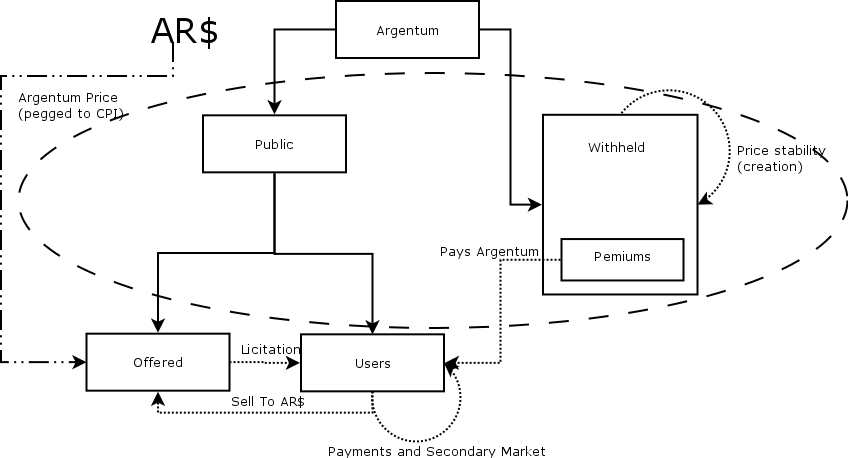}
    \caption{Expansion Model}
    \label{fig:expansion_model}
\end{figure}

In the case of an expansionary phase (Figure \ref{fig:expansion_model}), it is possible to find investment opportunities on the market, expanding the \textbf{Argentum}s owned by users. In such case, tokens will be created based on earnings and put on the withheld account. If opportunities are found, \textbf{Argentum} are generated and disponibilize to users. Also, with fixed periodicity the system is set to transfer  \textbf{Argentum} to pay for interest and premiums, while Users can transfer tokens between themselves.

\begin{figure}
    \centering
    \includegraphics[width=\linewidth]{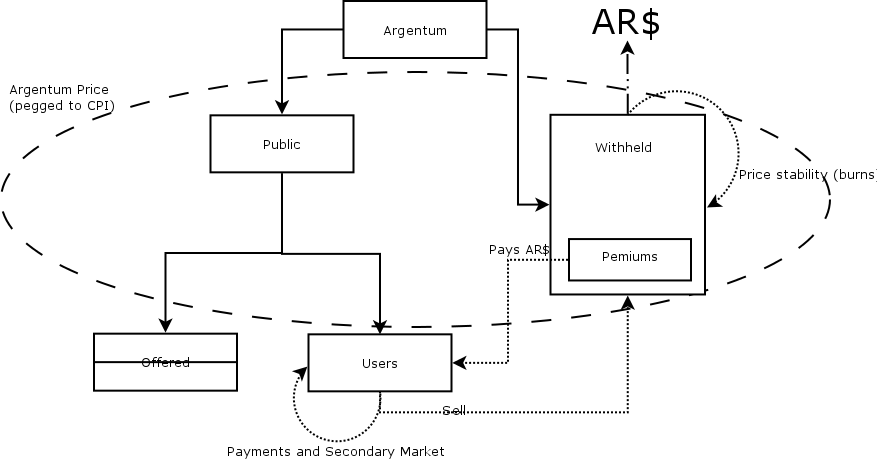}
    \caption{Contraction Model}
    \label{fig:contraction_model}
\end{figure}

In the case of a contractionary phase (Figure \ref{fig:contraction_model}), on the contrary, there is a lack of opportunities  to continue growth, so \textbf{Argentum} has to reduce the tokens owned by the Public \footnote{Never-the-less, this situation is not desirable and all the efforts of the projects has to be on generating investment opportunities. It has to be noted that there is a strategy in place to mitigate the damage that may be produced by contractory periods using an anticyclic fund and promoting the virtuous circle. No matter with that, the contractory phase is a mandatory process that has to be designed in order to keep system's credibility in different scenarios.}. If that occurs, tokens will be deactivated from the  withheld account to maintain valuation. As no opportunities are found, no amounts are transferred o the offered account, which remains in zero. However, with fixed periodicity the system is  set to transfer Argentine Pesos to pay for interest and premiums, while users can transfer tokens between themselves.

\subsection{Financial model}
\textbf{Argentum}'s sucess is tied up to the success of the financial team that ensures that the value deposited by \textbf{Argentum} holders is maintained over time. As explained in \ref{section:argentum_concept}, tokens will be created only if enough financial opportunities have been found. Those opportunities have to comply with the following criteria:

\begin{enumerate}
  \item \textbf{Low risk}: normally meaning that they will have a low return, but they have a very low risk of defaulting.
  \item \textbf{High liquidity}: that is, that they can be realized in a short period of time.
\end{enumerate}

It is interesting to notice that although the proyect aims to contribute to the estabilization of the economy, events outside of \textbf{Argentum} may occur, such as hyperinflation that may force the adoption of other rules in order to contain the damage. In such case, \textbf{Argentum Inc.} will try to maintain the value on the short term based on the opportunities available.

\subsection{User Model}
\textbf{Argentum} contemplates distint user roles in tension:
\begin{enumerate}
  \item \textbf{Saver}: He invests the excess excess amounts of his salary, normally in U\$D or fixed-term deposits, in order to save those amounts to spend them in the future. He expects - at least - to keep the purchase power over time, does not trust the banking system as he sees a viable option to keep cash in a safe, due to historical reason. He is under pressure from the authorities to put those amounts on the banking system, with negligible results. He is a cautious investor and sees the financial market as complex and unreliable. He deposits a small amount each month and transfers out occasionally.

  \item \textbf{Recurrent user}: He makes deposits in order to keep the purchasing power over the month, expects to make transactions directly from Argentum to other users and any excess stays on the account as savings. He will opt out some tokens to cover costs during the month, and he is not very interested on the long term investment so he is not competing for the premiums: he just expects easy of use and keep purchasing power over time of the remains of his operations. He deposits a greater amount each month but transfers out frequently.

  \item \textbf{Conservative Investor}: He invest significantly more than a saver, wants to keep the purchasing power over time and expects to get a moderate interest on his investment. He is willing to invest as long as it is easy and expects the interest and any premiums without any intervention. As his investment power is higher than the saver, he is able to access other investment options and prefers stability over maximizing return, since he is averse to risk. He deposits a lump sum and waits until maturity date.

  \item \textbf{Corporate Finance User}: He expects to lock the exchange rate for a number of months or years. He is very cautious about the result as a failure may jeopardize the project. However, he knows the market very well and looks for the best option to maintain the purchasing power in the long time.
\end{enumerate}

Although at first sight these user roles are in tension, it is expected to collaborate with each other on the system due to incentives alignment. The main difference between savers and conservative investors within the proposed project is the number of tokens they have in their accounts. The system is set to turn savers in investors over time.

\textbf{Argentum} is a project that takes very seriouly confidence and security, so only those that can legally own bank accounts on the targetted jurisdiction can be users of the system. Furthermore, as to conform to that philosophy, \textbf{Argentum} holds information details for each user to adopt applicable laws.

\subsection{Costs}
\textbf{Argentum} may charge a small fee for operating and recording costs.

Transactions are intended to apply in real time as long as the blockchain subsystem allows its registration. All transactions made between users of \textbf{Argentum} are irrestricted to the amount that the wallet holds, however, moving to other currencies from it may be restricted to the Reserve Requirement\cite{Wikipedia_reserve_requirement} of each jurisdiction per period of time with the cost of a transaction fee, to avoid bank runs\cite{Wikipedia_bank_run} which may affect negatively to the currency. If the user desires to move out to a greater percentage than said Reserve Requirement, he/she may have penalization in time or have the transaction costs increased substantially.

\subsection{Governance}
\textbf{Argentum} will be governed centrally by a corporation with dual incorporation, having its headquarters in a globally acepted market, such as United States or United Kindom and having a branch in Argentina to operate the token, named \textbf{Argentum Inc.}

The objectives of the proposed currency not covered by the previous section will be carried out by \textbf{Argentum inc.}, to cover the needs of a controlling entity that ensures their overall success, including the following areas:

\begin{enumerate}
  \item \textbf{Currency stabilization}: One of the goals of \textbf{Argentum} is to offer an instrument that has an stable value over time, in order to protect the purchasing power of token holders. Since there are many events may occur that can affect valuation, one of \textbf{Argentum inc.} core functions include monitor and correct any deviation from the targetted value.
  \item \textbf{Audit}: as part of the educational service of the project, the company is set to detect and take an action on "bad citizens" of the system, those who may use it in detriment of the majority. This area will greatly improve the credibility, stability and good name of the project in the long term.
  \item \textbf{Controversy Resolution and Legal support}: Inetably misunderstandings will happen and the legal team is set to resolve those situations without scalling them to other instances, on the base of \textit{bona fide} amongst token holders. Also, the legal team is set to handle relations with other entities that take part on this project and advices the other areas in legal matters.
  \item \textbf{Research and Development}: Evolve over time is a core principle of \textbf{Argentum inc.}, which will be reinvesting in new technologies and improvements to support its operations, not only creating new algorithms to detect financial opportunities but also providing the tools to developers who want to use this project as part of their own projects, as long as it does not interfere with \textbf{Argentum}'s own objectives.
\end{enumerate}

Due to the nature of a Blockchain operation model, in which operations can be carried out at any given time \textbf{Argentum Inc.} is intended to automate many of the operation areas with algorithms, ensuring fast and acurate response to users.

\subsection{Partners}
\textbf{Argentum} does not intend to rule out the banking institutions, for which they can continue their normal operations in a electronic based economy, offering better earnings on savings than Argentum, loans and storing the transactions made by their clients as a fail safe resource.

In fact, the \textbf{Argentum Inc.} is set to associate with financial institutions in order to acchieve faster results on an ever changing unstable economy.

In order to ensure fast results to entice users and a collaborative environment to operate, at least the following roles are meant to be covered by partner institutions:

\begin{enumerate}
  \item \textbf{Identity grantor}: a partner in charge of holding the correlation between the wallets and their owners registering vital data to map those wallets in the real world.
  \item \textbf{Retail}: a partner in charge of handling the relations with general consumers, registering and offering them tokens to purchase.
  \item \textbf{Corporate}: a partner in charge of handling the relations with corporate consumers, registering and offering them tokens to purchase.
  \item \textbf{Financial operator}: a partner in charge of handling the excution on the financial market, based on signals provided by the algorithm trading team.
\end{enumerate}

Partners will benefit from offering a new and enticing new product in the markets they operate, getting preferent rates and a portion of the earnings.

\subsection{Data Ecosystem}
As with any currency, the activity can lead to an inflation rate that - if not properly controlled - can lead ultimately to the lack of trust by their users and its demise. As such, as part of the analysis of this project, an algorithm has been created that finds opportunities with US Dollar futures, offered by ROFEX\cite{ROFEX_site}, yielding an impressive yield in the back testing simulations.

Nevertheless, it is important to remark that algorithms have a very high \textit{wear and tear}\cite{Wikipedia_wear_and_tear} cost, so an ongoing effort to maintain and improve algorithms will be needed. To that extent, the system is expected to provide a data analysis platform in which any user can download a data sample, develop an algorithm and get a reward based on the results. Under this schema, registered data analysts will be trained with sample data (which of course is of no use in the real world), and will be showcased to Argentum and other organizations to get better oportunities. Some experiences have been already implemented, most notably Numerai\cite{Numerai_webpage}.

This data model is intended to create a "data ecosystem" with the best talents in each economy, which will help to further ensure the ultimate success of the project. Whatismore, the success of the project may mean that better algorithms can be developed not only for Argentum but for other, riskier, projects.

\subsection{Implementation Stages}
It is clear that such a fundamental development cannot carried in short time or alone, as it will require the joint effort of numerous actors to be successfull. An outline can be seen in Table \ref{tab:general_implantation_stages}.

The main factor to consider is that -in order to make sure that it is stable over time- during the setup period \textbf{Argentum} will constitute a high risk fund with negative returns.

\begin{figure}
    \centering
    \includegraphics[width=\linewidth]{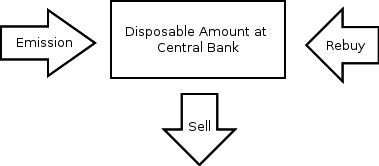}
    \caption{Disposable Amount}
    \label{fig:disposable_amount}
\end{figure}

Also, a deep monitoring of the \textbf{Argentum} in circulation will be conducted in order to make sure that the demand meets the Argentum offering using tools such as \textit{disponibilization} and \textit{rebuy}. In this context, rebuying will be done when users move out of \textbf{Argentum} and selling is done to banking institutions by means of depositing fiat currency as it can be seen in Figure \ref{fig:disposable_amount}.

As such, the final user, who buys \textbf{Argentum} from a banking institution, will ultimately experience two things: a currency that follows the inflation pattern, and a return made on fixed periods made by the financial team on his/her account.

% -*- coding: utf-8 -*-
% !Rnw root = ../argentum.Rnw
\section{Associated Products}

\textbf{Argentum} represents the core project, we consider that it is - by itself - not very useful without projects that use it as its foundation. In fact, since \textbf{Argentum}'s mission is to maintain the purchasing power  over time, it is the perfect tool to create a set of products for a more customized experience. For that reason, we envision that the implementation will go along with two associated projects, namely \textbf{Aurum} and \textbf{platinum}, being those projects the one that will drive \textbf{Argentum} to success.

% -*- coding: utf-8 -*-
% !Rnw root = ../argentum.Rnw

\subsection{Aurum}
\textbf{Aurum} is meant to provide a long term, low risk saving strategy for the working class, leveraging most of its functionality in \textbf{Argentum}.

Unlike \textbf{Argentum}, it is customized to guide its holders into saving, producing a premium based on performance on maturity date, while maintaining its value over time like \textbf{Argentum}.

\subsubsection{Key differences}
\begin{enumerate}
  \item \textbf{Social Aim}: The main motivation for this token is to provide a tool for middle class citizens of unstable economies to be able to harness the financial opportunities without a direct involvement in the market to allow them to maintain the purchasing power of their savings over time.

  \item \textbf{Rules}:
  \begin{enumerate}
\item \textbf{Saving Rule}: \textbf{Aurum} valuation is pegged to the Argentine Consumer Price Index (IPC es-CPI en)\cite{Wikipedia_IPC_Argentina}, which is the recommended  inflation proxy in the world\cite{cpi_ilo_manual,cpi_manual_updates} for household expenditures. So, \textbf{Argentum} will be a constant (deflected) valuation measure for Argentina's goods and services, which is considered useful in countries with high inflation\footnote{In Argentina, for example, real state and capital assets are commonly traded in $U\$D$.}. Along with this, \textbf{Aurum}  is intended to pay a moderate interest on the currency, around 2\%, which the user will see as an incoming transaction to his/her account.

  \item \textbf{Investment Rule}: Purchase of \textbf{Aurum} will be regulated by a monthly bidding process. Users who acquired tokens through the bidding process are eligible for a premium. For those cases, ten semesters after the purchase, users are entitled to receive up to a premium rate (around 70\%) of the net position of the proportional earnings amount.

  \item \textbf{Aurum Rule}: \textbf{Aurum}  monthly purchase price will be defined by a methodology explained in \ref{section:mensual_bidding} with the aim of penalizing speculation and concentration of purchases.  That is, those who wish to acquire larger amounts of \textbf{Aurum}, will receive a highest purchase price. This will discourage acquisitions of \textbf{Aurum}  for speculative proposes.

Following this rule, the objective proposed by \textbf{Aurum} is to provide advanced financial tools to common people, previously available only to high performance individuals and corporations, in order to promote saving habits.

  \end{enumerate}

  \item \textbf{Permanence bonus}: In order to motivate users to store their wealth in \textbf{Argentum}, a bonus price will be in place with the excess earnings resulting from the financial operation, as indicated in section \ref{section:argentum_concept}.

  \item \textbf{Secondary Market}: Although transactions amongst \textbf{Argentum} users are expected, they will not be encouraged in order to incentivate savings.
\end{enumerate}

% -*- coding: utf-8 -*-
% !Rnw root = ../argentum.Rnw

\subsubsection{Monthly bidding}
\label{section:mensual_bidding}

The mensual bidding is the process in which interested users are able to purchase coins at the lowest cost. The offer to the public is derived by the investment oportunities predicted and not related to the historic returns. The bidding will be a fair process to allow people to access to the coins in the same conditions. Both savers and investors access to the same bidding.

Following \textbf{Argentum}'s philosophy, which disponibilizes a financial tool for the average user, the bidding process inspired on the D'Hondt methodology\cite{Wikipedia_Dhondt_method}, named after Belgian mathematician Victor D'Hondt, who described it in 1878 for proportional allocation of parliamentary seats to the parties.

Initially, the monthly bidding can be described as follows, although it can be adjusted based on the experience and the market, a task that will be undertaken by \textbf{Argentum Inc.}

\begin{enumerate}
  \item \textbf{expression of interest}: Users deposits the compromised saving/investement ammount in $AR\$$ into their accounts strictly before the bidding time. investors must explicit a discounted rate for access to cut the process will be  implemented in 3 stages.

  The bidding will happen initially at the beggining of the second week. In such event, the user deposits an amount in the $AR\$$ account and sets the price he/she is willing to pay. The system then runs a \textit{deconcentrative bidding}, a process designed to give equal opportunities to both savers and investors.
    \item \textbf{bidding}: Users deposit the amounts to purchase the token and set the maximum price they are willing to accept.
    \item \textbf{execution}: The system decides how the \textbf{Argentum} will  be distributed based on behaviour (that is, if the user is a good platform citizen) and price. During this step, \textbf{Argentum Inc.} will charge for the coins distributed, refund the remaining amount and transfer them to the bidder's wallet to complete the operation.
\end{enumerate}

All premiums and interest payments are based on historic earnings and doesn't take into account the investment oportunities and are paid on \textbf{Argentum} (expansion model) or $AR\$$ (contraction model) on fixed periods to be determined by \textbf{Argentum Inc.}.

% -*- coding: utf-8 -*-
% !Rnw root = ../argentum.Rnw

\subsection{Platinum}
We that some holders will be less risk-adverse and thus a customized product can be created for them, which allow them to get a better return on investment, at the expense of the risk.

This project is key to \textbf{Argentum}, as it can leverage on the opportunities metodologies and algorithms that discovered to be are far too risky for  \textbf{Argentum} operation.

\subsubsection{Key differences}
\begin{enumerate}
  \item Although \textbf{Platinum} valuation may be pegged to the Argentine CPI, it will provide a return on investment based on tokens owned and their share on the profits.
  \item \textbf{Platinum} can be transferred between holders freely, and have the same selling restrictions as \textbf{Argentum}.
  \item \textbf{Platinum} uses all the infraestructure that is developed for \textbf{Argentum}, as long as its fit, including riskier algorithms and methodologies.
  \item The underlying fund can be deposited either in \textbf{Argentum} or other assets that may not be as liquid as in the case of the \textbf{Argentum}, allowing investment in riskier, long term assets.
\end{enumerate}

% -*- coding: utf-8 -*-
% !Rnw root = ../argentum.Rnw
\section{Benefits}
\label{A3CDF833-7291-4D84-9342-04E0A108DDA8}
\textbf{Argentum} is intended to:
\begin{enumerate}
  \item Mitigate transaction costs between the banking institutions
  \item Isolate citizens from the local economic instabilities, providing a nearly risk free economic tool to store wealth and save surplus earnings for future use, without the need of entering on the financial market.
  \item Create a base from which other projects can build upon, providing goods and services with the confidence that they will maintain the value over time.
  \item Help to create a paperless economy in a safe way, implemented on best practices.
\end{enumerate}
% -*- coding: utf-8 -*-
% !Rnw root = ../argentum.Rnw
\section{Conclusion}
Following the study, it is clear that there is a lack of a financial instrument to empower the middle class of unstable countries. In this order of ideas, Argentum may be a viable option to allow common citizens to save and spend without the worries of a major economic event that may damage their financial stability. As such, Argentum aims to:
\begin{enumerate}
  \item Mitigate transaction costs between the banking institutions
  \item Isolate citizens from the local economic instabilities, providing a risk free economic tool to store wealth and save surplus earnings for future use, without the need of entering on the financial market.
  \item Help to create a paperless economy in a safe way, implemented on best practices.
\end{enumerate}

With that in mind, it is feasible to indicate that \textbf{Argentum} may be a way to stabilize an economy in a fast and secure way, based on concepts that can be replicated nearly with no effort in different jurisdictions, while at the same time provides a tool to build upon other projects.

% -*- coding: utf-8 -*-
% !Rnw root = ../argentum.Rnw
\section{Anexes}
\subsection{History of Currency}
\subsubsection{Barter}

Barter is a system of exchange where goods or services are directly exchanged for other goods or services without using a medium of exchange, such as money\cite{barter_wikipedia}, and was the first system implemented by societies to acquire or sell in ancient times.
Although it may be sufficient for many small scale economies, it is extremely difficult to control and - although in use in some jurisdictions, normally as a measure of emergency - it tends to be unstable over time due to the fact that is nearly imposible to value one good based on any other. Some of its more clear limitations include:
\begin{enumerate}
	\item \textbf{Double coincidence of wants}. For barter to occur between two parties, both parties need to have what the other wants.
	\item \textbf{There is no common measure of value}. In a monetary economy, money plays the role of a measure of value of all goods, so their values can be assessed against each other; this role may be absent in a barter economy.
	\item \textbf{Indivisibility of certain goods}. If a person wants to buy a certain amount of another's goods, but only has for payment one indivisible unit of another good which is worth more than what the person wants to obtain, a barter transaction cannot occur.
	\item \textbf{Difficulty in storing wealth}. If a society relies exclusively on perishable goods, storing wealth for the future may be impractical.

\end{enumerate}

\subsubsection{Currency}
As a consequence the creation of a currency occurred. A currency (from Latin \textit{currens}) refers to money in any form when in actual use or circulation as a medium of exchange, especially circulating banknotes and coins, and in a broader sense to the system of money (monetary units) in common use in a jurisdiction.

Its history go up to 2,000 b.c., used first in Mesopotamia and then Egypt as a form of receipt representing grain stored in temple granaries. In this early stages, metals were used as symbols to represent value stored in the form of commodities. However, it had a fundamental flaw: in an era where it was not possible to store assets, the value of a medium could only be as sound as the forces that defended the store. To solve that, a number of treaties were stablished to allow merchants' safe passage but they were not enough to solve the issue. As a consequence, metal itself became the store of value being mined, weighed, and stamped into coins. Coins could be forged, but at least some guarantee was given to the taker that a certain amount of the metal was received, which lead to the creation of weight units for them and created a unit of value that, in turn, lead to the creation of banking.

During the Tang and Song Dynasties in China, banknotes were created as a means to relieve merchants from exchanging large amounts of coins, implemented later in numerous regions. They had some advantages (they reduced the need of transporting large amounts of money, it facilitated loans, enabled the sale of stock in joint-stock companies, and the redemption of those shares in paper) but they had also disadvantages: as the note has no intrinsic value, there were no constraint in printing more notes than they have specie to back them, which in turn created a new inflationary pressure. As a consequence, banknotes would often create inflationary bubbles which could collapse if people began demanding hard money, causing the demand for paper notes to fall to zero. However, it was also very addictive, as the speculative profits of trade and capital creation were so large.

This situation lead to the gold standard, by which every note printed had to be backed up by precious metals. Most of the industrialized world was in some form of it by 1900. By the end of the XX century all countries are in some form of of floating fiat currencies in the sense that the bank note is no longer backed up by precious metals anymore. Fiat Currency can be defined as an intrinsically valueless object or record that is widely accepted as a means of payment.

\subsubsection{Alternative Currency}
An alternative currency (or private currency) is any currency used as an alternative to the dominant national or multinational currency systems. They are created by an individual, corporation, or organization, they can be created by national, state, or local governments, or they can arise naturally as people begin to use a certain commodity as a currency. Mutual credit is a form of alternative currency, and thus any form of lending that does not go through the banking system can be considered a form of alternative currency.

\subsection{Crypto Currencies}
Crypto Currencies are a type of Alternative Coin, based on concepts like encryption and decentralized databases. Their most prominent exponents are BitCoin\cite{bitcoin_site}, Ethereum\cite{ethereum_site} (both based on the BlockChain), IOTA\cite{iota_site}, based on the Tangle, and the long awaited Libra \cite{Libra_site}.

As with every currency, a number of disadvantages are present, including the following, as indicated by the Stanford university\cite{Stanford_BitCoin_Disadvantages}:
\begin{enumerate}
	\item \textbf{No Buyer Protection}: When goods are bought using Crypto Currencies, and the seller doesn't send the promised goods, nothing can be done to reverse the transaction. This problem can be solved using a third party escrow service like ClearCoin, but then, escrow services would assume the role of banks, which would cause BitCoins to be similar to a more traditional currency.
	\item \textbf{No Valuation Guarantee}: Since there is no central authority governing the Crypto Currencies, no one can guarantee its minimum valuation. If a large group of merchants decide to "dump" any of them and leave the system, its valuation will decrease greatly which will immensely hurt users who have a large amount of wealth invested there. The decentralized nature of most Crypto Currencies is both a curse and blessing.
	\item \textbf{Valuation Fluctuation}: The value of most Crypto Currencies is constantly fluctuating according to demand. As of June 2nd 2011, one BitCoins was valued at \$9.9 on \href{https://www.coindesk.com}{coindesk.com} . It was valued to be less than \$1 just 6 months ago. This constant fluctuation will cause BitCoin accepting sites to continually change prices. It will also cause a lot of confusion if a refund for a product is being made. For example, if a t shirt was initially bought for 1.5 BTC, and returned a week later, should 1.5 BTC be returned, even though the valuation has gone up, or should the new amount (calculated according to current valuation) be sent? Which currency should BTC tied to when comparing valuation? These are still important questions that the BitCoin community still has no consensus over.
	\item \textbf{Built in Deflation}: This is specially harmful in BitCoins, since the total number is capped at 21 million, it will cause deflation. Each BitCoin will be worth more and more as the total number of BitCoins maxes out. This system is designed to reward early adopters. Since each BitCoin will be valued higher with each passing day, the question of when to spend becomes important. This might cause spending surges which will cause the BitCoin economy to fluctuate very rapidly, and unpredictably.
	      However, due to the lack of protection, they normally suffer from
	\item \textbf{Energy consumption}: In the case of BlockChain based Crypto Currencies, due to the process needed to produce a new hash, there is a huge energy consumption, which creates serious issues in most areas. For example, although it is expected to decline over time, BitCoin was consuming in July 2017 as much energy as Denmark\cite{Bitcoin_energy_Consumption_1}. In January 2018, it was set to consume more than Argentina\cite{Bitcoin_energy_Consumption_2}.
	\item \textbf{Storage needs}: Due to the characteristics of the BlockChain in which stores every single object in every instance of the database, its data storage needs are huge, creating a bottleneck in processing the storage for new objects. This, although is not a serious problem while the use of the Currency is marginal, over time will make storage of the full database nearly impossible to the average user, defeating the purpose of total transparency.
\end{enumerate}

\subsection{Currency Stability}
One of the main tools Governments have to deal with economic issues are collectively known as Monetary Policies, and they are normally carried out by the Central Bank or Currency Board of each Jurisdiction, often targeting an inflation rate or interest rate to ensure price stability and general trust in the currency. Generally speaking, the monetary policies can be:
\begin{enumerate}
	\item \textbf{Expansionary}: it happens when the economy is stimulated by maintaining short-term interest rates at a lower than usual rate or increasing the total supply of money in the economy more rapidly than usual,  It is traditionally used to try to combat unemployment in a recession by lowering interest rates in the hope that less expensive credit will entice businesses into expanding. This increases aggregate demand (the overall demand for all goods and services in an economy), which boosts short-term growth as measured by gross domestic product (GDP) growth. Expansionary monetary policy usually diminishes the value of the currency relative to other currencies (the exchange rate), thus having an inflationary effect.

	\item \textbf{Contractionary}: this case maintains short-term interest rates higher than usual or which slows the rate of growth in the money supply or even shrinks it. This slows short-term economic growth and lessens inflation. Contractionary monetary policy can lead to increased unemployment and depressed borrowing and spending by consumers and businesses, which can eventually result in an economic recession if implemented too vigorously.
\end{enumerate}

Nevertheless, during economic hardships Governments tend to fund their deficit by expanding the economy, which rapidly can out of control, depreciating the currency and seriously undermining its credibility in the market. When this happens, as the currency is unable to retain value over time, citizens need to find ways to store wealth, aggravating the situation, which can lead to what is called hyperinflation. Examples can be found in  Argentina\cite{wikipedia_argentina_riots_2001}, Rusia\cite{Russia_hyperinflation} and now Venezuela\cite{Venezuela_hyperinflation}.

\subsubsection{Stable Coins}
However, when thinking on stability coins on the Crypto world, it is possible to say that it is a cryptocurrency that is pegged to another stable asset, like gold or the U.S. dollar. It's a currency that is global, but is not tied to a central bank and has low volatility. This allows for practical usage of using cryptocurrency like paying for things every single day\cite{Forbes_explaining_stable_coins}.

In that sense, the most common crypto currencies, Bitcoin and Ethereum, are very volatile, which makes them inconvenient to use on the day to day basis. Price changes are a shock to the consumers and prevents them from using that coin as an exchange of value.

Ideally, an optimal stable cryptocurrency should have the following four traits: price stability, scalability, privacy, and decentralization\cite{Forbes_explaining_stable_coins}. However, out of the main options in the market today (Tether\cite{Tether_website}, MarkerDao\cite{MarketDao_website}, Havven\cite{Havven_website}), all of them are pegged to either the US Dollar or Euro.

\subsection{Argentina's Case}
Argentina's history with inflation is long and complex, leading to what is known as the \textit{Argentine Paradox}\cite{Wikipedia_Economic_history_of_argentina}. Although being a very resource rich country, its economic situation is highly tied to its political instability, dating back from the 1930s, before which it was  was one of the most stable and conservative countries until the Great Depression, after which it turned into one of the most unstable.

At the end of the 80s decade, an hyperinflation event arose, which seriously hindered the economy, forcing government to lock the exchange rate in what it was known as "convertibilidad". It lasted until 2002, which after the harsh economic crash of 2001, when it was replaced with a \textit{dirty floating} exchange rate which created a serious imbalance in the economy, which at this time, in 2018 is still affecting the day to day life of its residents. As of 2018, Argentina - while making great efforts to stabilize the economy are being made by the Goverment - remains unstable.

In Argentina there are a number of elements that threatens the economic and monetary stability, including:
  \begin{enumerate}
    \item Negative real interest rate.
    \item High Inflation levels, occasionally with hyperinflation\cite{EY_hiperinflationary_economies}.
    \item CPI Manipulation, with negative effects on the long run\cite{Wikipedia_indec, La_nacion_INDEC_the_lying_machine}.
    \item Recurrent devaluation events, with the consequent overshooting and speculation over the exchange rate equilibrium value\cite{Wikipedia_argentina_great_depression}.
    \item Recurrent defaults, the last one in 2001\cite{Economist_Argentinas_collapse}, resulting in a general lack of confidence on the Banking and Financial markets, among common users.
    \item Pendular changes in economic and monetary policies, including the Central Bank mission and role on the market, which creates instability\cite{The_currency_crisis_argentina}.
    %Cambios pendulares de política económica y monetaria, incluyendo la misión del Banco Central, con la consiguiente inestabilidad que esto genera \cite{Wikipedia_Curve_phillips}.
  \end{enumerate}

As a consequence, low risk investment instruments are not common which makes nearly unfeasible or too complex for the common user to maintain the purchasing power in the long run. This paper addresses the issues exposed with the aid of machine learning\cite{Wikipedia_machine_learning} techniques on the financial market and blockchain related technologies\cite{Wikipedia_blockchain} for the accounting and user side.

%Como corolario, las inversiones de bajo riesgo no son usuales, y por tanto no son opciones viables para el usuario medio. Este trabajo intenta dar una solución a esta problemática aplicando técnica de Inteligencia Artificial para gerenciamiento financiero (aplicacion de inversiones) y Blockchain para registro contable (ahorristas e inversores).

\subsection{Argentine Financial Landscape}
After a contraction of 1.8\% in 2016 and 2.9\% in 2017, Argentina now sees itself inmersed into a monetary crisis that required an International Monetary Fund program of  US\$50.000 Million in April to help maintain the current exchange rate with no success, which in turn rushed Argentines to rush for Dollars, since they are perceived as the way to mantain the purchasing value over time.

\printglossaries

\bibliographystyle{IEEEtran}
\bibliography{argentum}

\end{document}